\documentclass[12pt]{article}
\usepackage{epsfig}
\usepackage{hyperref}%

\date{}
\begin{document}

\newcommand{\beq}{\begin{equation}}
\newcommand{\eeq}{\end{equation}}
\newcommand{\nn}{\nonumber}

\def\ii{\'{\i}}
\def\r{\rightarrow}
\def\err{\end{array}}
\def\bea{\begin{eqnarray}}
\def\eea{\end{eqnarray}}
\def\bp{{\bf p}}
\def\bk{{\bf k}}
\def\bq{{\bf q}}
\def\ttau{\tilde{\tau}}
\def\tchi{\tilde{\chi}}
\def\trho{\tilde{\rho}}
\def\teps{\tilde{\epsilon}}
\def\tnu{\tilde{\nu}}
\def\tgamma{\tilde{\gamma}}
\def\G{G_{\mu\nu}}
\def\g{g_{\mu\nu}}
\def\E{{\mathcal{E}}_{\mu\nu}}
\def\P{{\mathcal{P}}_{\mu\nu}}
\def\h{h_{\mu\nu}}

\title{Dark Radiation Dynamics on the Brane}

\author{Rui Neves\footnote{E-mail: \tt rneves@ualg.pt}\hspace{0.2cm} and
Cenalo Vaz\footnote{E-mail: \tt cvaz@ualg.pt} \\
{\small \em \'Area Departamental de F{\ii}sica/CENTRA, FCT,
Universidade do Algarve}\\
{\small \em Campus de Gambelas, 8000-117 Faro, Portugal}
}

\maketitle

\begin{abstract}
We investigate the dynamics of a spherically symmetric vaccum on a
Randall and Sundrum 3-brane world. Under certain natural conditions, the
effective Einstein equations on the brane form a closed system for
spherically symmetric dark radiation. We determine exact
dynamical and inhomogeneous solutions, which are shown to depend on
the brane cosmological constant, on the dark radiation tidal charge
and on its initial energy configuration. We identify the
conditions defining these solutions as singular or as globally regular.
Finally, we discuss the confinement of gravity to the vicinity of the
brane and show that a phase transition to a regime where gravity is
not bound to the brane may occur at short distances during the
collapse of positive dark energy density on a realistic de Sitter
brane.
\vspace{0.5cm}

\end{abstract}

\section{Introduction}

It is possible to localize matter fields in a 3-brane world embedded in a higher
dimensional space if gravity is allowed to propagate away from the brane in the
extra dimensions (see for example \cite{CLED} and the early related work in
\cite{LED}). The extra dimensions were usually compactified to a finite volume but
it was realized that they could also be infinite and non-compact as in the model of
Randall and Sundrum (RS) who proposed gravity to be confined to the vicinity of
the brane by the warp of a single extra dimension \cite{RS}.

In the RS brane world scenario the observable universe is a 3-brane boundary of
a non-compact $Z_2$ symmetric 5-dimensional Anti-de Sitter (AdS) space. The
matter fields are restricted to the brane but gravity exists in the whole AdS
bulk. The classical dynamics is defined by the 5-dimensional Einstein field
equations with a negative bulk cosmological constant and a Dirac delta source
representing the brane. The original RS solution involves a non-factorizable
bulk metric for which the light-cone into the fifth dimension is closed by
an exponential warp factor \cite{RS}

\beq
d{\tilde{s}^2}={e^{-2\gamma|y|}}{\eta_{\mu\nu}}d{x^\mu}d{x^\nu}+d{y^2},
\eeq
where $\eta_{\mu\nu}$ is the Minkowski metric in 4 dimensions, $\gamma=\sqrt{
-\tilde{\Lambda}{\tilde{\kappa}^2}/6}$ with $\tilde{\Lambda}$ the negative bulk
cosmological constant and
$\tilde{\kappa}^2= 8\pi/{\tilde{M}_{\mbox{\footnotesize
p}}^3}$ with $\tilde{M}_{\mbox{\footnotesize p}}$ denoting the
fundamental 5-dimensional Planck mass. The brane cosmological constant
was assumed to be fine-tuned to zero and so
$\tilde{\Lambda}=-{\tilde{\kappa}^2} {\lambda^2}/6$
or $\gamma={\tilde{\kappa}^2}\lambda/6$ where $\lambda$ is the brane
tension. Weak field perturbations around this static vaccum solution
\cite{RS}-\cite{GKR} were also analysed to reveal a new way to solve
the hierarchy problem
and show that 4-dimensional Einstein gravity is effectively
recovered on the brane at low energy scales if the AdS radius
$1/2\gamma$ is small enough. In particular Newton's
potential $V_N$ generated by a mass $M$ receives warp corrections
which to leading order give

\beq
{V_N}={{{G_N}M}\over{r}}\left(1+{1\over{{\gamma^2}{r^2}}}\right),\label{Npld}
\eeq
where $G_N$ is Newton's gravitational constant. Extensions of this
scenario were rapidly developed to allow for example non-fine-tuned
branes \cite{SMS,KR} and thick branes \cite{CEHS}. The Gauss-Codazzi formulation
used in \cite{SMS} was discussed earlier in \cite{CGC}.

Friedmann-Robertson-Walker cosmologies \cite{BDL}-\cite{SD}, static
black holes \cite{CHR}-\cite{ND} and stars \cite{GM}, the Vaidya
solution \cite{DG} as well as Oppenheimer-Snyder gravitational collapse
solutions \cite{BGM,GD} were also discovered within the RS brane world
scenario. However, it should be noted that the static black hole
solutions discussed in \cite{EHM} are for a 4-dimensional bulk and the
existence of black hole solutions in the bulk that reduce to the
a static black hole localized on the brane remain unknown.
Moreover, cosmological perturbations have
started to be implemented \cite{COSp}-\cite{LMSW}. So far a consistent
adjustment to the available data coming out of astrophysics, cosmology
and also high energy particle colision experiments has been shown to
hold \cite{RM}. In addition, it has been shown that the AdS/CFT
duality \cite{CFT} is compatible with the RS scenario \cite{RSCFT}.

Studies of the gravitational collapse of matter have been limited to static or
homogeneous dynamical models. Notable examples are the 5-dimensional
black string solution \cite{CHR}, the tidal Reissner-Nordstr\"om black
hole \cite{DMPR}, the Vaidya solution on the brane \cite{DG} and the
Oppenheimer-Snyder model \cite{BGM,GD}. As a result it has become clear
that the exterior vaccum of a collapsing distribution of
matter on the brane cannot be static \cite{BGM,GD}. This was shown to
be a consequence of the presence of gravitational modes
in the bulk, which act as a kind of intrinsically dynamical dark
matter in the whole brane world. These gravitational degrees of
freedom are generated by the existence of Weyl curvature in the
bulk and are a source of inhomogeneities in matter clouds located on
the brane.

The purpose of this paper is to analyse the spherically symmetric,
inhomogeneous vacuum on the brane. This is a necessary step towards
the understanding of a realistic collapse problem in the RS
scenario. We take an effective 4-dimensional point of view following the
Gauss-Codazzi covariant geometric approach
\cite{SMS,CGC,BDL} (see also Ref. \cite{RM} for a review and
notation). In section 2 we introduce the Einstein vaccum
field equations on the brane. Although these equations do not, in
general, form a closed system, additional conditions may be imposed to
form a closed system described by two free parameters, {\it viz.,} the
brane cosmological constant $\Lambda$ and the dark radiation
tidal charge $Q$. In this section we also present a discussion on the
localization of gravity near the brane. For small but positive
$\Lambda$ we show that while for $Q<0$ the tidal acceleration
in the off-brane direction is always negative, implying acceleration
towards the brane, for $Q>0$ the acceleration changes sign at a critical
short distance. In section 3 we present  $\Lambda$ and $Q$ dependent
inhomogeneous dynamical solutions. We show that they take the
LeMa\^{\i}tre-Tolman-Bondi form and depend on the initial
configuration of the dark radiation, which is parametrized by a single
function, $f=f(r)>-1$. This is interpreted as the energy function. The
marginally bound solution ($f(r)=0$) is the static, zero mass tidal
Reissner-Nordstr\"om black hole solution of \cite{DMPR}. In
section 4 we analyse the dynamics of the non-marginally bound,
inhomogeneous solutions and characterize precisely
how they depend on $\Lambda$, $Q$ and $f(r)$. We also identify the
conditions under which the dark radiation dynamics leads to the
formation of a singularity as the final outcome of collapse or to a
globally regular evolution with a bounce developing
inside the vaccum. We conclude in section 5.

\section{Vaccum Field Equations on the Brane}

In the RS brane world scenario the Einstein field equations in the
bulk are \cite{RS,RM}

\beq
{\tilde{G}_{AB}}={\tilde{\kappa}^2}\left[-\tilde{\Lambda}{\tilde{g}_{AB}}+
\delta(y)\left(-\lambda{g_{AB}}+{T_{AB}}\right)\right],\label{5Dfe}
\eeq
The 4-dimensional 3-brane is located at $y=0$ and is a fixed point of the
$Z_2$ symmetry. Its
induced metric is ${g_{AB}}={\tilde{g}_{AB}}-{n_A}{n_B}$
where $n_A$ is the spacelike unit normal to the brane. Matter confined
to the brane is characterized by the energy-momentum tensor $T_{AB}$
and
satisfies ${T_{AB}}{n^B}=0$. The 5-dimensional metric is of the RS form

\beq
d{\tilde{s}^2}={\tilde{g}_{AB}}d{x^A}d{x^B}=d{y^2}+{g_{\mu\nu}}
d{x^\mu}d{x^\nu},
\eeq
where $g_{\mu\nu}$ should display a dependence on the fifth
dimension which localizes gravity near the brane and closes the light-cone.

According to the effective geometric approach
\cite{SMS,CGC,BDL,RM} the induced Einstein field equations on the brane are
obtained from Eq. (\ref{5Dfe}), the Gauss-Codazzi equations and the
Israel conditions
with $Z_2$ symmetry. In the vaccum $T_{AB}$ is set to zero, as a
consequence of which the induced equations take the form

\beq
\G=-\Lambda\g-\E,\label{efe}
\eeq
where

\beq
\Lambda={{\tilde{\kappa}^2}\over{2}}
\left(\tilde{\Lambda}+
{{{\tilde{\kappa}^2}{\lambda^2}}\over{6}}\right)
\eeq
is the brane cosmological constant and $\E$ is the
limit on the brane of the projected 5-dimensional Weyl tensor,

\beq
\E={\lim_{y\to 0\pm}}{\delta^A_\mu}{\delta^B_\nu}{\mathcal{E}_{AB}}=
{\lim_{y\to 0\pm}}{\delta^A_\mu}{\delta^B_\nu}{\tilde{C}_{ACBD}}{n^C}{n^D}.
\eeq
It is a symmetric and traceless
tensor due to the Weyl symmetries and is constrained by the
conservation equations

\beq
{\nabla_\mu}{{\mathcal{E}}^\mu_\nu}=0\label{ce},
\eeq
obtained from Eq. (\ref{efe}) as a result of the Bianchi identities.

The system
of vaccum field equations on the brane defined by Eqs. (\ref{efe})
and (\ref{ce}) may naturally be interpreted as defining the interaction of
4-dimensional Einstein gravity with matter represented by the
traceless energy-momentum tensor \cite{DMPR}

\beq
{\kappa^2}{T_{\mu\nu}}=-\E,\label{emt}
\eeq
where ${\kappa^2}=8\pi/{M_{\mbox{\footnotesize p}}^2}$ with
$M_{\mbox{\footnotesize p}}=1/\sqrt{G_N}$ the
effective Planck mass on the brane and \cite{SMS}

\beq
{\kappa^2}={{\lambda{\tilde{\kappa}^4}}\over{6}}.\label{contps}
\eeq
Using general algebraic symmetry properties it is possible to
write $\E$ as \cite{RM4dp}

\beq
\E=-{{\left({{\tilde{\kappa}}\over{\kappa}}\right)}^4}\left[{\mathcal{U}}
\left({u_\mu}{u_\nu}+{1\over{3}}\h\right)+\P+{{\mathcal{Q}}_\mu}{u_\nu}+
{{\mathcal{Q}}_\nu}{u_\mu}\right],
\eeq
where $u_\mu$ such that ${u^\mu}{u_\mu}=-1$ is the 4-velocity field
and $\h=\g+{u_\mu}{u_\nu}$ is the tensor which projects
orthogonaly to
$u_\mu$. The forms $\mathcal{U}$, $\P$ and ${\mathcal{Q}}_\mu$
represent the effects on the brane of the free gravitational
field in the bulk. Thus, $\mathcal{U}$ is the
effective energy density,

\beq
\mathcal{U}=-{{\left({{\kappa}\over{\tilde{\kappa}}}\right)}^4}\E{u^\mu}{u^\nu},\eeq
$\P$ is the anisotropic stress,

\beq
\P=-{{\left({{\kappa}\over{\tilde{\kappa}}}\right)}^4}\left[{1\over{2}}\left({h_\mu^\alpha}
{h_\nu^\beta}+{h_\nu^\alpha}
{h_\mu^\beta}\right)-{1\over{3}}\h{h^{\alpha\beta}}\right]{\mathcal{E}_{\alpha\beta}},
\eeq
and ${\mathcal{Q}}_\mu$ is the effective
energy flux

\beq
{{\mathcal{Q}}_\mu}={{\left({{\kappa}\over{\tilde{\kappa}}}\right)}^4}{h_\mu^\alpha}
{\mathcal{E}_{\alpha\beta}}{u^\beta}.
\eeq
Even though $\E$ is trace free and has to satisfy the conservation equations,
in general it cannot be fully determined on the brane. This is to be expected
because observers confined to the brane cannot predict all bulk effects without
a knowledge of the solution of the full 5-dimensional Einstein equations. As a
consequence, the effective 4-dimensional geometric theory is not closed. However,
under additional simplifying assumptions about the bulk degrees of freedom it
is possible to close the system if $\E$ is constrained in such a way that it
may be completely determined by its symmetries and by the conservation equations.
This is what happens when $\P=0$ \cite{RM4dp} or if $\P\not=0$ when a spherically
symmetric brane is static \cite{DMPR}.

Let us now show that it is possible to take a non-static spherically symmetric brane with
$\P\not=0$ and still close the system of dynamical equations. Consider the general,
spherically symmetric metric in comoving coordinates $(t,r,\theta,\phi)$,

\beq
d{s^2}=\g
d{x^\mu}d{x^\nu}=-{e^\sigma}d{t^2}+A^2d{r^2}+{R^2}d{\Omega^2},
\label{met}
\eeq
where $d{\Omega^2}=d{\theta^2}+{\sin^2}
\theta d{\phi^2}$, $\sigma=\sigma(t,r)$, $A=A(t,r)$, $R=R(t,r)$ and $R$ is interpreted as the
physical spacetime radius. If there is no net energy flux, then ${{\mathcal{Q}}_\mu}=0$. Again,
if the stress is isotropic, ${\mathcal P}_{\mu\nu}$ will have the general form

\beq
\P={\mathcal{P}}\left({r_\mu}{r_\nu}-{1\over{3}}\h\right),
\eeq
where ${\mathcal{P}}={\mathcal{P}}(t,r)$ and $r_\mu$ is the unit radial vector, given in the
above metric by ${r_\mu}=(0,A,0,0)$. The projected Weyl tensor then takes the diagonal
form

\beq
{{\mathcal{E}}_\mu^\nu}={{\left({{\tilde{\kappa}}\over{\kappa}}\right)}^4}\mbox{diag}\left(\rho,-
{p_r},-{p_T},-{p_T}\right),
\eeq
where the energy density and pressures are respectively $\rho={\mathcal{U}}$,
${p_r}=(1/3)\left({\mathcal{U}}+2{\mathcal{P}}\right)$ and ${p_T}=(1/3)\left({\mathcal{U}}-{\mathcal{P}}
\right)$. Substituting in the conservation Eq. (\ref{ce}) we obtain the following expanded
system \cite{TPS}

\[
2\frac{\dot A}{A}\left(\rho+{p_r}\right)=-2\dot{\rho}-
4{{\dot{R}}\over{R}}\left(\rho+{p_T}\right),
\]
\beq
\sigma'\left(\rho+{p_r}\right)=-{p_r}'+4{{R'}\over{R}}({p_T}-{p_r}),\label{cee}
\eeq
where the dot and the prime denote, respectively, derivatives
with respect to $t$ and $r$. A synchronous solution is obtained by taking the equation of state
$\rho+{p_r}=0$, giving ${\mathcal{P}}=-2{\mathcal{U}}$. Then the conservation Eqs. (\ref{cee})
separate and we obtain

\beq
\dot{{\mathcal{U}}}+4{\dot{R}\over{R}}{\mathcal{U}}=0={\mathcal{U}}'+
4{R'\over{R}}{\mathcal{U}}.
\eeq
Consequently the effective
energy density has the dark radiation form \cite{DR}

\beq
{\mathcal{U}}={{\left({{\kappa}\over{\tilde{\kappa}}}\right)}^4}
{Q\over{R^4}},\label{KfB}
\eeq
where the dark radiation tidal charge $Q$ is constant. Thus for the
gravitational dynamics of spherically symmetric but still inhomogeneous
dark radiation $\E$ is fully determined to be

\beq
\E=-{Q\over{R^4}}\left({u_\mu}{u_\nu}-
2{r_\mu}{r_\nu}+\h\right).\label{ten}
\eeq
Note that this projected Weyl tensor $\E$ is a simple dynamical
extension of the
corresponding form associated with the static tidal
Reissner-Nordstr\"om solution
on the brane \cite{DMPR}. In spite of the existing pressures in the
dark radiation setting we may safely take the synchronous comoving
frame, for which $\sigma=0$. Then the intricate general field equations
\cite{TPS} simplify and substituting Eq. (\ref{ten}) in
Eq. (\ref{efe}) we obtain

\beq
\G=-\Lambda\g+{Q\over{R^4}}\left({u_\mu}{u_\nu}-
2{r_\mu}{r_\nu}+\h\right),\label{dem}
\eeq
a consistent and exactly solvable closed system for the two unknown
functions, $A(t,r)$ and $R(t,r)$.

Clearly, the dark radiation dynamics depends on $\Lambda$ and
$Q$ \cite{BGM,GD}. An important point to note is that these parameters
have a direct influence
on the localization of gravity in the vicinity of the brane
(see also Ref. \cite{SD}). To understand how consider the tidal
acceleration away from the brane as measured by
brane observers \cite{RM4dp}. For the dark radiation vaccum we are
considering such acceleration is \cite{DMPR}

\beq
-{\lim_{y\to 0\pm}}{\tilde{R}_{ABCD}}{n^A}{\tilde{u}^B}{n^C}{\tilde{u}^D}=
{{\tilde{\kappa}^2}\over{6}}\tilde{\Lambda}+{Q\over{R^4}},
\eeq
where $\tilde{u}_A$ is the
extension off the brane of the 4-velocity field which satisfies
${\tilde{u}^A}{n_A}=0$ and ${\tilde{u}^A}{\tilde{u}_A}=-1$. For the
gravitational field to be
localized near the brane the tidal acceleration must be negative. The
condition for this to happen is

\beq
{\tilde{\Lambda}}{R^4}<-{{6Q}\over{\tilde{\kappa}^2}}.
\eeq
Consequently, the localization of gravity for all $R$ is only possible
if $\tilde{\Lambda}<0$ and $Q\leq 0$ or $\tilde{\Lambda}=0$ and $Q<0$.
In terms of brane parameters this implies that $\Lambda<{\Lambda_c}$
with ${\Lambda_c}=
{\tilde{\kappa}^4}{\lambda^2}/12$ and
$Q\leq 0$ or $\Lambda={\Lambda_c}$ and $Q<0$. For
$\Lambda<{\Lambda_c}$ and $Q>0$ the
gravitational field will only remain localized near the 
brane if $R>{R_c}$ where ${R_c^4}=3Q/({\Lambda_c}-\Lambda)$. On the
other hand for $\Lambda>{\Lambda_c}$ and
$Q<0$ confinement is restricted to the epochs $R<{R_c}$. If
$\Lambda\geq{\Lambda_c}$ and $Q\geq 0$ then
gravity is always free to propagate far away from the brane.

Since the dark radiation term comes from the electric or Coulomb
part of the 5-dimensional Weyl tensor \cite{SMS} the geometry of the
bulk may assign to $Q$ any real value. Note however that in our
settings the weak, strong and dominant energy conditions \cite{RW} all
independently imply $Q\geq
0$ because $\rho={\mathcal{U}}$, ${p_r}=-{\mathcal{U}}$ and
${p_T}={\mathcal{U}}$. This is consistent with other
studies of dark radiation effects \cite{DR} but not with perturbation theory
and the negative induced energy condition \cite{ND} which need $Q<0$
\cite{SMS,DMPR,ND}. Current observations do not yet
constrain the sign of $Q$ \cite{DMPR,LMSW,RM,qexp}. So in what follows
we will keep $Q$ as real parameter. The same does not hold for the cosmological
constant $\Lambda$. Indeed, according to the present data
$\Lambda\sim{10^{-84}}{\mbox{GeV}^2}$ \cite{EXP}. On the other hand
${\tilde{M}_{\mbox{\footnotesize p}}}>{10^8}\mbox{GeV}$ and
${M_{\mbox{\footnotesize p}}}\sim{10^{19}}\mbox{GeV}$ imply by
Eq. (\ref{contps})
$\lambda>{10^8}{\mbox{GeV}^4}$ \cite{LMSW}. Since using once again
Eq. (\ref{contps})
we may write ${\Lambda_c}={\kappa^2}\lambda/2$ then ${\Lambda_c}$ has
a lower limit, ${\Lambda_c}>{10^{-29}}{\mbox{GeV}^2}$. Hence according
to observations $\Lambda$ must be positive and below the critical
value $\Lambda_c$, $0<\Lambda<{\Lambda_c}$. This implies a bulk with
negative cosmological constant, $\tilde{\Lambda}<0$. Note as well that
the same
conclusion holds if $M_{\mbox{\footnotesize
      p}}$ is in the TeV range as $\Lambda_c$ increases when
$M_{\mbox{\footnotesize p}}$ decreases. Finally note that
when $0<\Lambda<{\Lambda_c}$ and $Q>0$ then gravity is not confined
near the brane for $R\leq{R_c}$. This is a non-perturbative result
which is consistent with the energy conditions \cite{RW} for dark
radiation on the brane.
So far experiments have not been able to detect
any confinement phase transition to a high energy regime where gravity
is not bound to the brane. If
such a transition does occur at $1/{R_c}\sim 1\mbox{TeV}$ then for
${\Lambda_c}\sim{10^{-29}}{\mbox{GeV}^2}$ we get
$Q\sim{10^{-42}}{\mbox{GeV}^{-2}}$, a value well below current
astrophysical or weak gravity experimental constraints on $Q$
\cite{DMPR,LMSW,RM,qexp}.

For completeness in what
follows we will also discuss the de Sitter (dS) models with
$\Lambda\geq{\Lambda_c}$ which correspond to a dS bulk, the brane AdS
setting where $\Lambda<0$ and the brane with $\Lambda=0$.

\section{Dynamical Inhomogeneous Solutions}

To completly define the brane world dark radiation vaccum dynamics we
need to solve Eq. (\ref{dem}). The solutions determine the evolution
of the brane world as a whole and so are cosmological in nature. In
the proper frame $\sigma=0$ and so the remaining components of the
spherically symmetric metric given in Eq. (\ref{met}) must satisfy the
off-diagonal equation

\beq
{G_{tr}}={2\over{AR}}\left(\dot{A}R'-\dot{R}'A\right)=0,
\eeq
which implies

\beq
A={{R'}\over{H}},\label{AfB}
\eeq
where $H=H(r)$ is an arbitrary positive function of $r$. Then
introducing Eq. (\ref{AfB}) in the trace equation

\beq
-{G_t^t}+{G_r^r}+2{G_\theta^\theta}=-{2\over{AR}}
\left(\ddot{A}R+2\ddot{R}A\right)=
-2\Lambda+{{2Q}\over{R^4}}\label{Tfe}
\eeq
and integrating in $r$ leads to

\beq
\ddot{R}={\Lambda\over{3}}R+{Q\over{R^3}}.
\eeq
Note that at this point we have chosen to set the arbitrary
integration function (of time) to zero. We make this choice here because such a
function would correspond to an initial dust mass. A further integration gives

\beq
\dot{R}^2={\Lambda\over{3}}{R^2}-{Q\over{R^2}}+f,
\label{deq}
\eeq
where $f=f(r)$ is an arbitrary function of $r$. This function is naturally
interpreted as the energy inside a shell labelled by $r$ in the dark
radiation vaccum. Integrating Eq. (\ref{deq}) we obtain

\beq
\pm t+\psi=\int{{RdR}\over{\sqrt{{\Lambda\over{3}}{R^4}+f{R^2}-Q}}},
\label{dieq}
\eeq
where $\psi=\psi(r)$ is another arbitrary function of $r$ and the signs
$+$ or $-$ refer respectively to expansion or collapse.

As it stands the dark radiation is characterized by three
arbitrary functions of $r$, namely, $H$, $f$ and $\psi$. However, we
have the freedom to rescale the radial coordinate $r$. Then we may
impose on the initial hypersurface $t=0$ the condition

\beq
R(0,r)=r\label{rsc}
\eeq
and so prescribe $\psi$ to be given by

\beq
\psi=\int{{RdR}\over{\sqrt{{\Lambda\over{3}}{R^4}+f{R^2}
-Q}}},\label{vpsi}
\eeq
where the r.h.s is evaluated at $t=0$. Note that condition (\ref{rsc}) also defines the initial dark radiation
effective density profile. Using the initial distribution of
velocities we can also determine the energy function $f$. $H$ is then
determined by the remaining Einstein equation,

\beq
{G_r^r}=-\Lambda-{Q\over{R^4}},
\eeq
as $H=\sqrt{1+f}$ and consequently the metric takes the
LeMa\^{\i}tre-Tolman-Bondi form

\beq
d{s^2}=-d{t^2}+{{{R'}^2}\over{1+f}}d{r^2}+{R^2}d{\Omega^2},
\eeq
where $f>-1$.

Note that for the marginally bound models corresponding to $f=0$ this
metric describes a static solution. Indeed performing a transformation
from the Tolman-Bondi
coordinates $(t,r)$ to the curvature coordinates $(T,R)$ such that

\beq
T=t+\int dR {{R\sqrt{-Q+{\Lambda\over{3}}{R^4}}}\over{{\Lambda\over{3}}{R^4}-{R^2}-Q}},
\eeq
we find

\beq
d{s^2}=-\left(1+{Q\over{R^2}}-{\Lambda\over{3}}{R^2}\right)d{T^2}+
{{\left(1+{Q\over{R^2}}-{\Lambda\over{3}}{R^2}\right)}^{-1}}d{R^2}+
{R^2}d{\Omega^2},
\eeq
which is the inhomogeneous static exterior of a collapsing sphere of
homogeneous dark radiation \cite{BGM} (see also Ref. \cite{GD}).
When $\Lambda=0$ this corresponds to the zero mass limit of the tidal
Reissner-Nordstr\"om black hole solution on the brane
\cite{DMPR}. The single black hole horizon only exists for $Q<0$ and
is located at ${R_h}=\sqrt{-Q}$. Consequently for $Q>0$ the
singularity at ${R_s}=0$ is naked. For non-zero $\Lambda$ the solution
has horizons at \cite{BGM}

\beq
{R_h^{\pm}}=\sqrt{{3\over{2\Lambda}}\left(1\pm\sqrt{1+{{4Q\Lambda}\over{3}}}\right)}.
\eeq
If $\Lambda>0$ and $Q<0$ then we have an inner horizon $R_h^-$ and an outer
horizon $R_h^+$. The two horizons merge for $Q=-3/(4\Lambda)$ and for
$Q<-3/(4\Lambda)$ the singularity at ${R_s}=0$ becomes naked. For
$\Lambda>0$ and $Q>0$ there is a single horizon at $R_h^+$. If
$\Lambda<0$ and $Q>0$ the singularity at ${R_s}=0$ is again naked. For
$\Lambda<0$ and $Q<0$ this no longer happens as a single horizon forms
at $R_h^-$.

In the absence of dark radiation we simply find the homogeneous dS or
AdS spaces \cite{HE}. In particular for $\Lambda>0$ the solution
takes the Friedmann-Robertson-Walker steady state form where
the spatial curvature $k$ is zero,

\beq
R=r\exp\left(\pm\sqrt{{\Lambda}\over{3}}t\right).
\eeq
Then it represents (half) homogeneous dS space \cite{HE}. The epoch
$R=0$ is regular but fixed as it is impossible to leave it classically.

For $\Lambda=0$ the marginally bound models allow us to calculate
the leading order contribution to Newton's potential.
Following Ref. \cite{DMPR} and keeping the coordinates $T$ and $R$ we
may introduce a mass $M$ for the
gravitational source to find the tidal Reissner-Nordstr\"om metric

\beq
d{s^2}=-\left(1-{{2{G_N}M}\over{R}}+{Q\over{R^2}}\right)d{T^2}+
{{\left(1-{{2{G_N}M}\over{R}}+{Q\over{R^2}}\right)}^{-1}}d{R^2}+
{R^2}d{\Omega^2}.
\eeq
Then Newton's potential is given by

\beq
{V_N}={{{G_N}M}\over{R}}-{Q\over{2{R^2}}}+{\mathcal{O}}
\left({1\over{R^3}}\right).
\eeq
Contrary to Eq. (\ref{Npld}) this is a short distance perturbative
expansion \cite{DMPR}. So just
as in the static limit our dynamical solutions with non-zero $f$
should be more adequate to the high energy regime as long as quantum
gravity effects can be neglected. The potential shows that
perturbatively it should be $Q\leq 0$ if anti-gravity effects are to
be avoided at short distances (see also Ref. \cite{ND}). Consequently
if a positive $Q$ is
to be allowed as is by the standard energy conditions \cite{RW} than
it must be seen as an intrinsically non-perturbative effect.

Let us now consider $f\not=0$. Depending on $f$ and on $\Lambda$
and $Q$ we distinguish several possible dynamical inhomogeneous solutions.

\subsection{dS dynamics: $\Lambda>0$}

In this setting Eq. (\ref{dieq}) can be cast in the form

\beq
\pm t+\psi={1\over{2}}\sqrt{{3\over{\Lambda}}}
\int{{dY}\over{\sqrt{Y^2-\beta}}},\label{dieq1}
\eeq
where $Y={R^2}+3f/(2\Lambda)$ and $\beta=\beta(r)$ is given by

\beq
\beta={3\over{\Lambda}}\left[{{3{f^2}}\over{4\Lambda}}+Q\right].
\eeq
Direct evaluation of this integral determines $R$ and shows that the
solutions are organized by $\beta$. For $\beta>0$ we find

\beq
\left|{R^2}+{{3f}\over{2\Lambda}}\right|=\sqrt{\beta}\cosh\left[\pm 2
\sqrt{{\Lambda\over{3}}}t+{\cosh^{-1}}\left(
{{\left|{r^2}+{{3f}\over{2\Lambda}}\right|}
\over{\sqrt{\beta}}}\right)\right],\label{sol1}
\eeq
where for $Q>0$ the energy function may span all its range,
$f>-1$,
but for $Q<0$ it must further satisfy $|f|>2\sqrt{-Q\Lambda/3}$. If
$\beta<0$ we obtain

\beq
{R^2}+{{3f}\over{2\Lambda}}=\sqrt{-\beta}\sinh\left[\pm 2
\sqrt{{\Lambda\over{3}}}t+{\sinh^{-1}}
\left({{{r^2}+{{3f}\over{2\Lambda}}}
\over{\sqrt{-\beta}}}\right)\right],
\eeq
where $Q<0$ and $|f|<2\sqrt{-Q\Lambda/3}$. For $\beta=0$ we get

\beq
\left|{R^2}+{{3f}\over{2\Lambda}}\right|=\left|{r^2}+{{3f}\over{2\Lambda}}
\right|\exp\left(\pm 2\sqrt{{\Lambda\over{3}}}t\right),\label{sol2}
\eeq
where $Q<0$ and $f=\pm 2\sqrt{-Q\Lambda/3}$.

Clearly, these cosmological solutions are explicitly dependent on
$r$ in a way which cannot be evaded by any coordinate
transformation. As such they imply an
intrinsically inhomogeneous dark radiation efective energy
density. Indeed, $R$ is not a factorizable function, a fact preventing a
reduction to the standard homogeneous dS or Robertson-Walker spaces \cite{HE}.

\subsection{AdS dynamics: $\Lambda<0$}

In this scenario Eq. (\ref{dieq}) can be written as

\beq
\pm t+\psi={1\over{2}}\sqrt{-{3\over{\Lambda}}}
\int{{dY}\over{\sqrt{\beta-Y^2}}}.
\eeq
Now the only possibility is $\beta>0$. Integrating we obtain
the solutions

\beq
{R^2}+{{3f}\over{2\Lambda}}=\sqrt{\beta}\sin\left[\pm 2
\sqrt{-{\Lambda\over{3}}}t+{\sin^{-1}}\left({{{r^2}+{{3f}\over{2\Lambda}}}
\over{\sqrt{\beta}}}\right)\right],\label{sol3}
\eeq
where for $Q<0$ the energy function can take any value in the range
$f>-1$ but for $Q>0$ it must also be in the interval
$|f|>2\sqrt{-Q\Lambda/3}$.

\subsection{Absence of cosmological constant: $\Lambda=0$}

In the limit $\Lambda=0$ Eq. (\ref{dieq}) is written as

\beq
\pm t+\psi=\int{{RdR}\over{\sqrt{f{R^2}-Q}}}.
\eeq
Then the solutions for $f\not=0$ are given by

\beq
{R^2}={1\over{f}}\left[Q+{{\left(\pm
        ft+\sqrt{f{r^2}-Q}\right)}^2}
\right],
\eeq
where $f{R^2}-Q>0$. Note that if $Q>0$ then $f>0$.

\subsection{Absence of dark radiation: $Q=0$}

When dark radiation is not present in the vaccum we have $Q=0$ and
then the brane world dynamics depends on $\Lambda$ and $f$ as can be seen by
going back to Eq. (\ref{deq}). Integrating it we obtain

\beq
\pm t +\psi(r)=\int{{dR}\over{\sqrt{{\Lambda\over{3}}{R^2}+f}}}.
\eeq
Let us first consider $\Lambda>0$. The solutions are now organized by
$f$. For $f>0$ we find

\beq
R=\sqrt{{3f}\over{\Lambda}}\sinh\left[\pm\sqrt{{\Lambda}\over{3}}t+
{\sinh^{-1}}\left(\sqrt{{\Lambda}\over{3f}}r\right)\right].
\eeq
If $-1<f<0$ we have

\beq
R=\sqrt{-{{3f}\over{\Lambda}}}\cosh\left[\pm\sqrt{{\Lambda}\over{3}}t+
{\cosh^{-1}}\left(\sqrt{-{{\Lambda}\over{3f}}}r\right)\right].\label{sdS0}
\eeq
Note that it is only for $f=-\Lambda{r^2}/3$ that solution (\ref{sdS0}) takes
the homogeneous dS form \cite{HE}.

When $\Lambda<0$ the relevant
integral is only defined for $f>0$ and the corresponding solution is

\beq
R=\sqrt{-{3f}\over{\Lambda}}\sin\left[\pm\sqrt{-{\Lambda}\over{3}}t+
{\sin^{-1}}\left(\sqrt{-{\Lambda}\over{3f}}r\right)\right].
\eeq
In this case the homogeneous AdS space solution \cite{HE} is again obtained
with the choice $f=-\Lambda{r^2}/3$.

\section{Physical Singularities and Regular Rebounces}

Both shell focusing singularities ($R=0$) and shell crossing
singularities ($R'=0$) are found depending on the dynamics prescribed
by Eq. (\ref{deq}). Shell crossing singularities are generally not
considered to be ``real'' singularities, so we will concentrate always
on shell focusing singularities in the following.

To determine the inhomogeneous dynamical solutions corresponding to
dark radiation gravitation on the brane we have
implicitly assumed the validity of the following regularity condition

\beq
V=V(R,r)={{\Lambda}\over{3}}{R^4}+f{R^2}-Q>0.\label{Regc}
\eeq
If Eq. (\ref{Regc}) is verified for all $R\geq 0$ then collapsing
dark radiation shells will meet at the physical shell focusing singularity
$R={R_s}=0$. This is what happens in the gravitational collapse of ordinary
dust matter when $\Lambda$ is set to zero \cite{PSJ}. However, if
$\Lambda\not=0$ the evolution curve of a collapsing dust cloud
has special rebouncing points at $R>0$ for which the dust potential is
zero \cite{DJCJ}. These
points may exist both when a singularity forms at ${R_s}=0$ and when it
does not. In the latter case the solution is globally regular and the
gravitational evolution of the dust cloud involves a collapsing phase
followed by continuous expansion after velocity reversal at the
rebounce point.

Not
surprisingly, an analogously rich structure of solutions can also be
uncovered in the gravitational dynamics of vaccum dark
radiation on the brane (see also the discussion in Ref. \cite{BGM}).
We start by writing
Eq. (\ref{deq}) in the following way

\beq
{R^2}{\dot{R}^2}=V.
\eeq
Starting from an initial (collapsing) state, a rebounce will occur for
a given shell whenever $\dot{R}=0$ before that shell becomes
singular. This happens when $V(R,r)=0$. Introducing $Y$ and $\beta$ we write the potential $V$ as the
following simple quadratic polynomial

\beq
V={{\Lambda}\over{3}}\left({Y^2}-\beta\right)
\eeq
and it becomes clear that no more than two regular rebounce epochs can be
found. Depending on $f$ and on the
values of $\Lambda$ and $Q$ none, one or two such epochs may be
associated with globally regular solutions.

\subsection{dS dynamics: $\Lambda>0$}

The dS dynamics corresponding to
$\Lambda>0$ is characterized by a concave phase space curve because

\beq
{{{d^2}V}\over{d{Y^2}}}={{2\Lambda}\over{3}}>0.
\eeq
This implies the existence of a phase of continuous expansion towards
infinity with ever increasing speed. Other phases are possible
depending on the parameters.

\begin{figure}[t]
\setlength{\unitlength}{1cm}
\begin{minipage}[t]{7.5cm}

\begin{picture}(7.5,5.0)
\epsfig{file=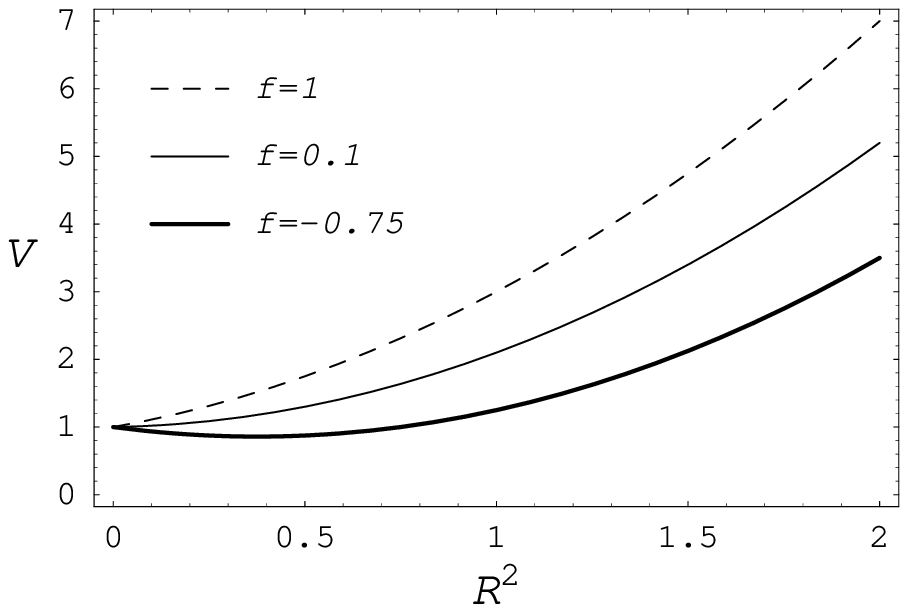,width=\hsize}
%\epsfxsize=1.0\hsize
%\epsfbox{plotd32a.eps}
%\label{fig:d32a}
\end{picture}\par
%\caption{Plots}
\end{minipage}
\hfill
\begin{minipage}[t]{7.5cm}
\begin{picture}(7.5,5.0)
\epsfig{file=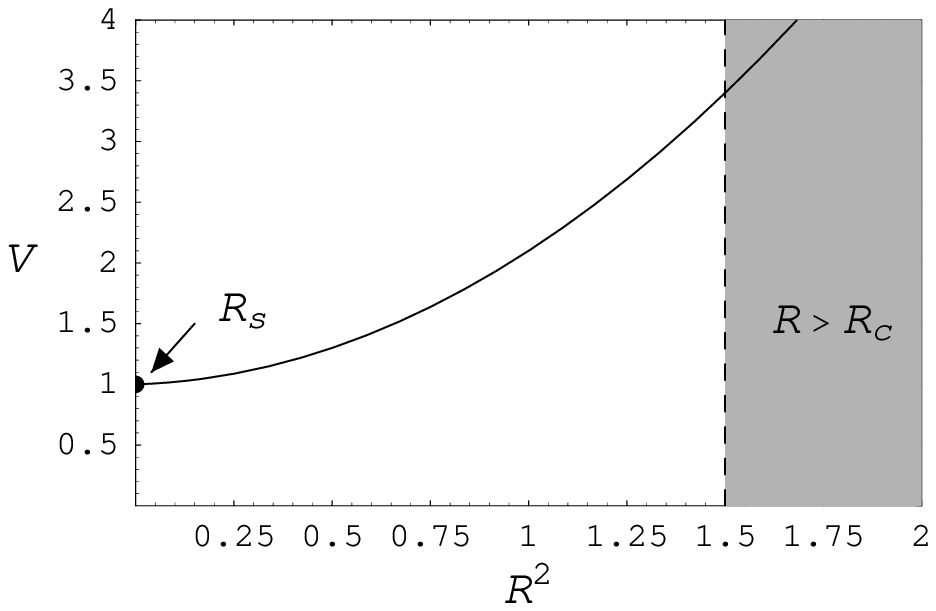,width=\hsize}
%\epsfxsize=\hsize
%\epsfbox{plotd32a.eps}
%\label{fig:d32a}
\end{picture}\par
%\caption{Plot}
\end{minipage}
\caption{\small Plots of $V$ for
$\Lambda>0$, $\beta<0$ and $Q<0$. Non-zero values of $f$ belong to the
interval $f>-1\bigwedge|f|<2\sqrt{-Q\Lambda/3}$ and correspond to
shells of
constant $r$. In the lefthand plot $\Lambda\leq{\Lambda_c}$ and
gravity is always bound to the brane. In the
righthand plot $\Lambda>{\Lambda_c}$ and the shaded region indicates
where gravity is not confined near the brane.}
\label{fig:Vpnn}
\end{figure}

For $\beta<0$ we have $Q<0$ and
$|f|<2\sqrt{-Q\Lambda/3}$. Clearly, $V$ is
positive for all $R\geq 0$. It satisfies $V(0,r)=-Q$ and grows to infinity with $R$ as
$\Lambda{R^4}$ (see Fig.~\ref{fig:Vpnn}). The dark radiation shells may either expand
continuously with ever increasing kinetic energy or collapse to a
shell focusing singularity at ${R_s}=0$ after a proper time
$t={t_s}(r)$ given by

\beq
{t_s}(r)={1\over{2}}\sqrt{{3\over{\Lambda}}}\left[{\sinh^{-1}}
\left({{{r^2}+{{3f}\over{2\Lambda}}}
\over{\sqrt{-\beta}}}\right)-{\sinh^{-1}}\left({{3f}
\over{2\Lambda\sqrt{-\beta}}}\right)\right].
\eeq
For $\Lambda\leq{\Lambda_c}$ gravity is always confined near the brane
during the cycle of evolution. However, if $\Lambda>{\Lambda_c}$ then
the gravitational field only stays confined for $R<{R_c}$. As a
consequence an expanding dark radiation vaccum goes through a phase
transition at $R={R_c}$ where the gravitational field ceases to be
confined to the brane. As we have seen this case and the large value
$\Lambda={\Lambda_c}$ are currently ruled out by observations.

\begin{figure}[t]
\setlength{\unitlength}{1cm}
\begin{minipage}[t]{7.5cm}

\begin{picture}(7.5,5.0)
\epsfig{file=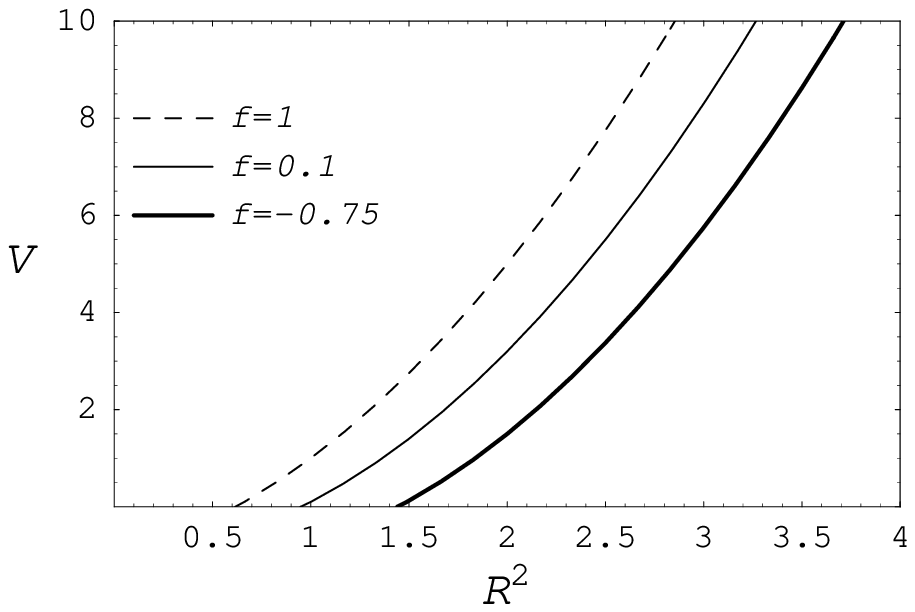,width=\hsize}
%\epsfxsize=1.0\hsize
%\epsfbox{plotd32a.eps}
%\label{fig:d32a}
\end{picture}\par
%\caption{Left}
\end{minipage}
\hfill
\begin{minipage}[t]{7.5cm}
\begin{picture}(7.5,5.0)
\epsfig{file=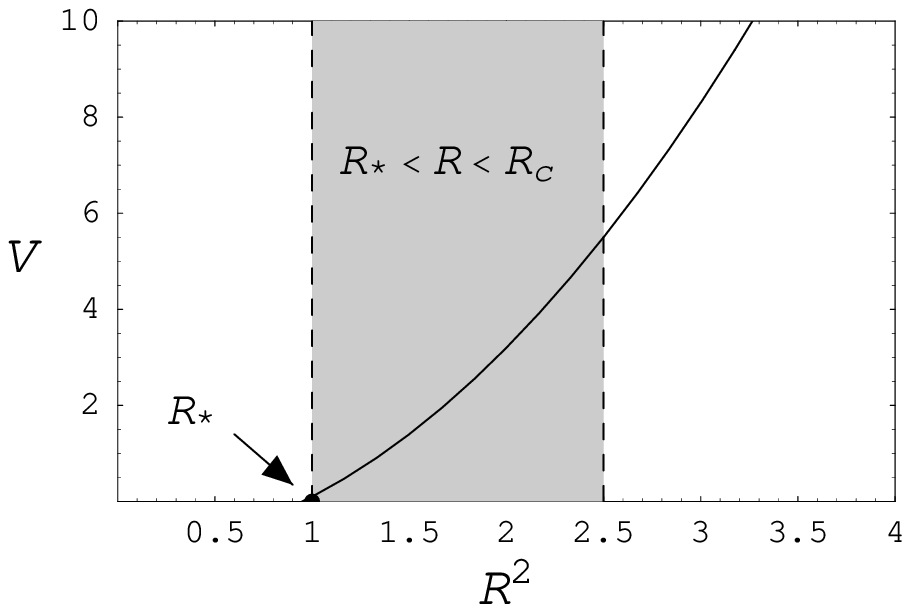,width=\hsize}
%\epsfxsize=1.0\hsize
%\epsfbox{plotd32a.eps}
%\label{fig:d32a}
\end{picture}\par
%\caption{Right}
\end{minipage}
\caption{\small Plots of $V$ for
$\Lambda>0$, $\beta>0$ and $Q>0$. Non-zero values of $f$ belong to the
interval $f>-1$ and correspond to shells of
constant $r$. In the lefthand plot $\Lambda\geq{\Lambda_c}$ gravity is
always free to move away from the brane. In the
righthand plot $\Lambda<{\Lambda_c}$ and the shaded region indicates
where gravity is not bound to the brane.}
\label{fig:Vppp}
\end{figure}

For $\beta>0$, $Q>0$ and $f>-1$ (see Fig.~\ref{fig:Vppp}) we find globally
regular solutions with a single rebounce epoch at $R={R_*}$ where

\beq
{R_*^2}=-{{3f}\over{2\Lambda}}+\sqrt{\beta}.
\eeq
This is the minimum possible radius a collapsing dark radiation shell
can have. The collapsing shells reach this epoch at the proper time
$t={t_*}(r)$ where

\beq
{t_*}(r)={1\over{2}}\sqrt{{3\over{\Lambda}}}{\cosh^{-1}}\left(
{{{r^2}+{{3f}\over{2\Lambda}}}
\over{\sqrt{\beta}}}\right).
\eeq
At this point they reverse their motion and expand forever with ever
increasing rate. The phase
space of
allowed dynamics defined by $V$ and $R$ is thus restricted to the
region $R\geq{R_*}$. Below
$R_*$ there is a forbidden region where the potential $V$ is negative.
In particular $V(0,r)=-Q<0$ which means that the
singularity at ${R_s}=0$ does not form and so the solutions are globally
regular. Note that if gravity is to
stay confined to the vicinity of the brane for $R>{R_*}$ then
$\Lambda<{\Lambda_c}$ and ${R_*}>{R_c}$. If not then
there is a phase transition epoch $R={R_c}$ such that for $R\leq{R_c}$
the gravitational field is no longer confined to the brane.
For $\Lambda\geq{\Lambda_c}$ gravity will always be free to propagate
far away from the brane a situation that as we have seen is not
allowed by current observations.

\begin{figure}[t]
\setlength{\unitlength}{1cm}
\begin{minipage}[t]{7.5cm}

\begin{picture}(7.5,5.0)
\epsfig{file=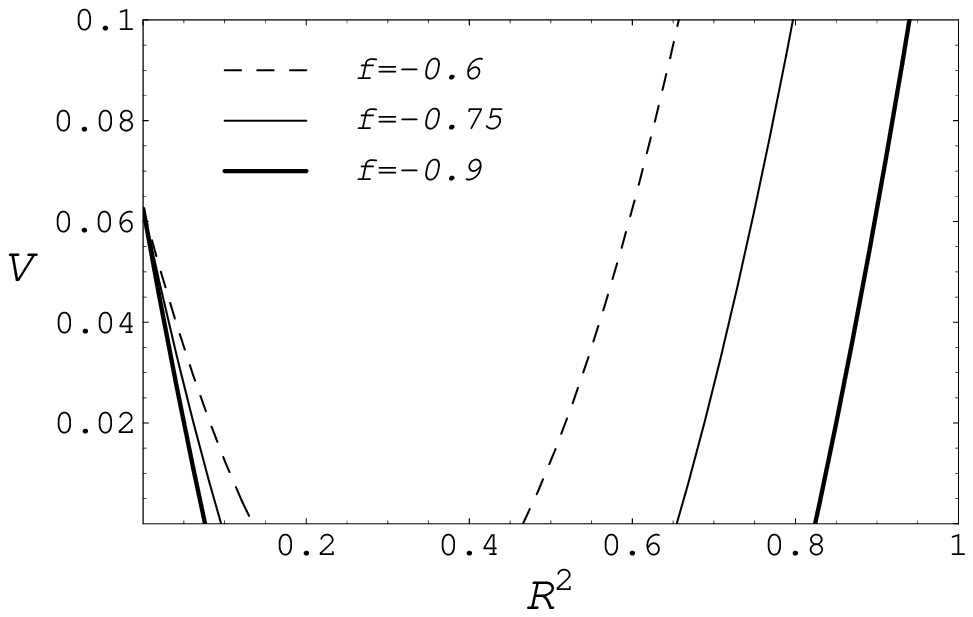,width=\hsize}
%\epsfxsize=1.0\hsize
%\epsfbox{plotd32a.eps}
%\label{fig:d32a}
\end{picture}\par
%\caption{Left}
\end{minipage}
\hfill
\begin{minipage}[t]{7.5cm}
\begin{picture}(7.5,5.0)
\epsfig{file=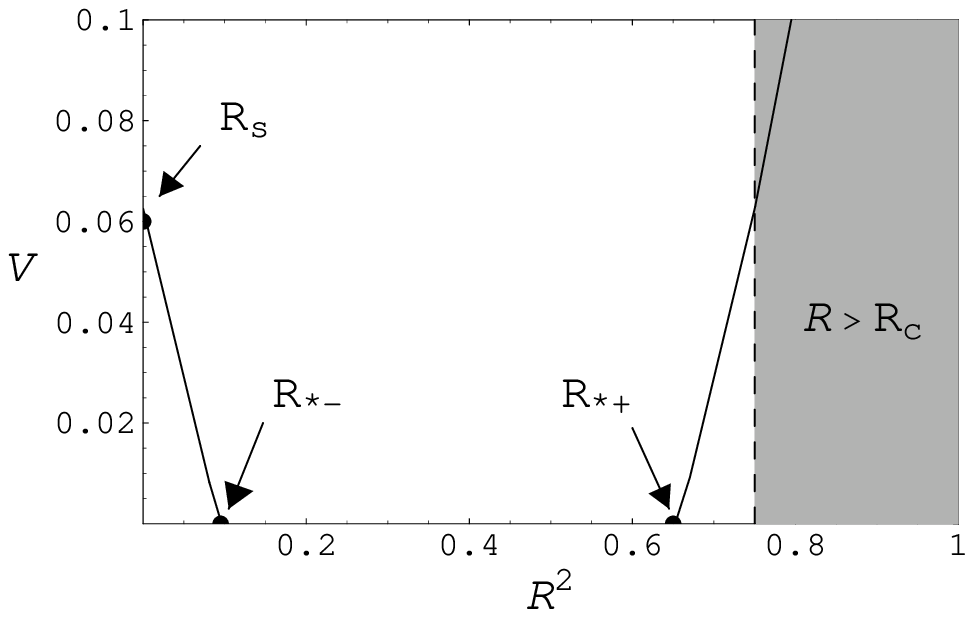,width=\hsize}
%\epsfxsize=1.0\hsize
%\epsfbox{plotd32a.eps}
%\label{fig:d32a}
\end{picture}\par
%\caption{Right}
\end{minipage}
\caption{\small Plots of $V$ for
$\Lambda>0$, $\beta>0$ and $Q<0$. Non-zero values of $f$ belong to the
interval $-1<f<-2\sqrt{-Q\Lambda/3}$ and correspond to
shells of constant $r$. In the lefthand plot
$\Lambda\leq{\Lambda_c}$ and gravity is
always confined near the brane. In the
righthand plot $\Lambda>{\Lambda_c}$ and the shaded region indicates
where gravity is not bound to the brane.}
\label{fig:Vppn}
\end{figure}

If $\beta>0$ and $Q<0$ we have seen that $f>-1$ has to
further satisfy
$|f|>2\sqrt{-Q\Lambda/3}$. If $f>2\sqrt{-Q\Lambda/3}$ there no
rebounce points in the allowed dynamical region $R\geq 0$ and the
evolution follows the case $\beta<0$ ilustrated in
Fig.~\ref{fig:Vpnn}. The
time
to reach the singularity at ${R_s}=0$ is now given by

\beq
{t_s}(r)={1\over{2}}\sqrt{{3\over{\Lambda}}}\left[{\cosh^{-1}}
\left({{{r^2}+{{3f}\over{2\Lambda}}}
\over{\sqrt{\beta}}}\right)-{\cosh^{-1}}\left({{3f}
\over{2\Lambda\sqrt{\beta}}}\right)\right].
\eeq
On the
other hand for $-1<f<-2\sqrt{-Q\Lambda/3}$ (see Fig.~\ref{fig:Vppn})
there are two rebounce
epochs at $R=R_{*\pm}$ with

\beq
{R_{*\pm}^2}=-{{3f}\over{2\Lambda}}\pm\sqrt{\beta}.
\eeq
Since $V(0,r)=-Q>0$ a singularity also forms at ${R_s}=0$. The
region between the two rebounce points
is forbidden because there $V$ is negative. The phase space
of allowed dynamics is thus divided in two disconnected
regions separated by the forbidden interval ${R_{*-}}<R<{R_{*+}}$. For
$0\leq R\leq{R_{*-}}$ the
dark radiation shells may expand to a maximum radius $R={R_{*-}}$ in
the time ${t_{*-}}={t_*}$ where

\beq
{t_*}(r)={1\over{2}}\sqrt{{3\over{\Lambda}}}{\cosh^{-1}}\left(
{{\left|{r^2}+{{3f}\over{2\Lambda}}\right|}
\over{\sqrt{\beta}}}\right).
\eeq
At this rebounce epoch the shells start to fall to the singularity
which is met after the proper time

\beq
{t_s}(r)={1\over{2}}\sqrt{{3\over{\Lambda}}}{\cosh^{-1}}
\left({{3|f|}\over{\Lambda\sqrt{\beta}}}\right).
\eeq
If $R\geq{R_{*+}}$ then there
is a collapsing phase to the minimum radius $R={R_{*+}}$ taking the
time ${t_{*+}}={t_*}$ followed by reversal
and subsequent accelerated continuous expansion. The singularity at
${R_s}=0$ does not form and so the solutions are globally
regular. Once more note that for $\Lambda\leq{\Lambda_c}$ gravity is
always confined near the brane
but for $\Lambda>{\Lambda_c}$ there is a phase transition epoch at $R={R_c}$
such that for $R>{R_c}$ gravity is free to propagate out of the
brane.
Only if ${R_{*-}}<{R_c}<{R_{*+}}$ this phase transition will not
occur. For $R<{R_{*-}}$ gravity is
confined near the brane and for $R>{R_{*+}}$ it will not.

\begin{figure}[t]
\setlength{\unitlength}{1cm}
\begin{minipage}[t]{7.5cm}

\begin{picture}(7.5,5.0)
\epsfig{file=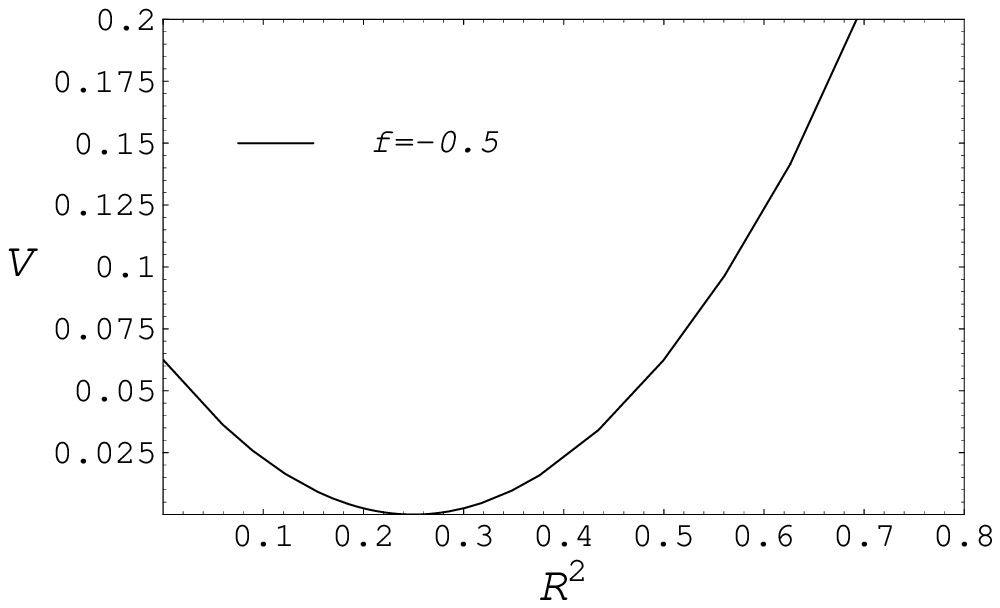,width=\hsize}
%\epsfxsize=1.0\hsize
%\epsfbox{plotd32a.eps}
%\label{fig:d32a}
\end{picture}\par
%\caption{Left}
\end{minipage}
\hfill
\begin{minipage}[t]{7.5cm}
\begin{picture}(7.5,5.0)
\epsfig{file=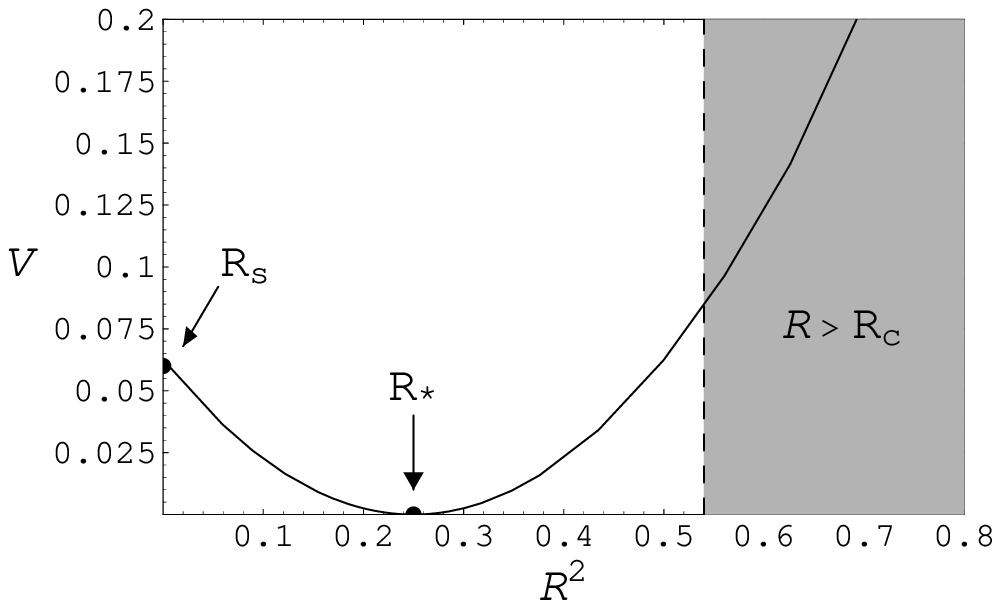,width=\hsize}
%\epsfxsize=1.0\hsize
%\epsfbox{plotd32a.eps}
%\label{fig:d32a}
\end{picture}\par
%\caption{Right}
\end{minipage}
\caption{\small Plots of $V$ for
$\Lambda>0$, $\beta=0$ and $Q<0$. Here $f=-2\sqrt{-Q\Lambda/3}$. In
the lefthand plot $\Lambda\leq{\Lambda_c}$ and gravity is
always confined near the brane. In the
righthand plot $\Lambda>{\Lambda_c}$ and the shaded region indicates
where gravity is not bound to the brane.}
\label{fig:Vp0n}
\end{figure}

If $\beta=0$ then again $Q<0$. For $f=-2\sqrt{-Q\Lambda/3}$
(see Fig.~\ref{fig:Vp0n}) the single
candidate to be a rebounce point is $R={R_*}$ with

\beq
{R_*}=\sqrt{-{{3Q}\over{\Lambda}}}.
\eeq
In this case $V(0,r)=-Q>0$ and then a singularity also forms at
${R_s}=0$. There is no forbidden region in phase space but the point
at $R_*$ turns out to be a regular fixed point which divides two
distinct dynamical regions. Indeed if a shell starts at $R={R_*}$
then it will not move for all times. If initially $R<{R_*}$ then
either the shell expands towards $R_*$ or it collapses to
the singularity. The time to meet the singularity is finite,

\beq
{t_s}(r)={3\over{2\Lambda}}\ln\left({{\sqrt{-{{3Q}\over{\Lambda}}}}\over{\left|
{r^2}-\sqrt{-{{3Q}\over{\Lambda}}}\right|}}\right),
\eeq
but the time to expand to ${R_*}$ is infinite. If initially $R>{R_*}$
then the collapsing dark radiation shells also take an infinite time to
reach $R_*$. If $f=2\sqrt{-Q\Lambda/3}$ there are no
real rebounce epochs and the collapsing dark radiation simply falls to the
singularity at ${R_s}=0$. The colision proper time is

\beq
{t_s}(r)=-{3\over{2\Lambda}}\ln\left({{\sqrt{-{{3Q}\over{\Lambda}}}}
\over{{r^2}+\sqrt{-{{3Q}\over{\Lambda}}}}}\right).
\eeq
Of course for $\Lambda\leq{\Lambda_c}$ then the gravitational field is
always confined to the brane but for
$\Lambda>{\Lambda_c}$ there is a phase transition epoch $R={R_c}$ in
either of the two dynamical regions such
that for $R>{R_c}$ gravity may freely move far into the bulk.

\subsection{AdS dynamics: $\Lambda<0$}

The AdS dynamics corresponding to
$\Lambda<0$ is characterized by a convex phase space curve because

\beq
{{{d^2}V}\over{d{Y^2}}}={{2\Lambda}\over{3}}<0,
\eeq
which does not allow a phase of continuous expansion to infinity.
Also in this scenario the only possibility is $\beta>0$.

\begin{figure}[t]
\center{\psfig{file=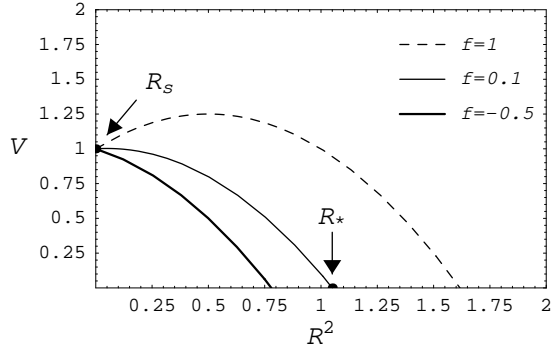,width=0.5\hsize}}
%\epsfxsize=1.0\hsize
%\epsfbox{plotd32a.eps}
\caption{\small Plots of $V$ for
$\Lambda<0$, $\beta>0$ and $Q<0$. Non-zero values of $f$ belong to the
interval $f>-1$ and correspond to shells of
constant $r$. Here gravity is
always bound to the vicinity of the brane.}
\label{fig:Vnpn}
\end{figure}

If $Q<0$ then we
may consider $f>-1$ (see Fig.~\ref{fig:Vnpn}). Just like the
corresponding dS case there
is a single rebounce point at $R={R_*}$ with

\beq
{R_*^2}=-{{3f}\over{2\Lambda}}+\sqrt{\beta}.
\eeq
Due to the convexity of the phase space evolution curve this is the
maximum radius which an expanding dark radiation shell can
reach. The time for this to happen is

\beq
{t_*}(r)={1\over{2}}\sqrt{-{3\over{\Lambda}}}\left[{\pi\over{2}}-{\sin^{-1}}
\left({{{r^2}+{{3f}\over{2\Lambda}}}
\over{\sqrt{\beta}}}\right)\right].
\eeq
Because $V(0,r)=-Q>0$ a singularity also forms at
${R_s}=0$. Consequently, the reversal at
$R={R_*}$ will be followed by a collapsing phase to the
singularity in the time

\beq
{t_s}(r)={1\over{2}}\sqrt{-{3\over{\Lambda}}}\left[{\pi\over{2}}-{\sin^{-1}}
\left({{3f}\over{2\Lambda\sqrt{\beta}}}\right)\right].
\eeq
This is in contrast with the dS scenario where
there was no formation of a singularity at ${R_s}=0$ and after reversal at
the rebounce epoch the dark radiation shells expand forever. Also,
because $\Lambda<{\Lambda_c}$ gravity remains confined near the brane
for all times.

\begin{figure}[t]
\setlength{\unitlength}{1cm}
\begin{minipage}[t]{7.5cm}

\begin{picture}(7.5,5.0)
\epsfig{file=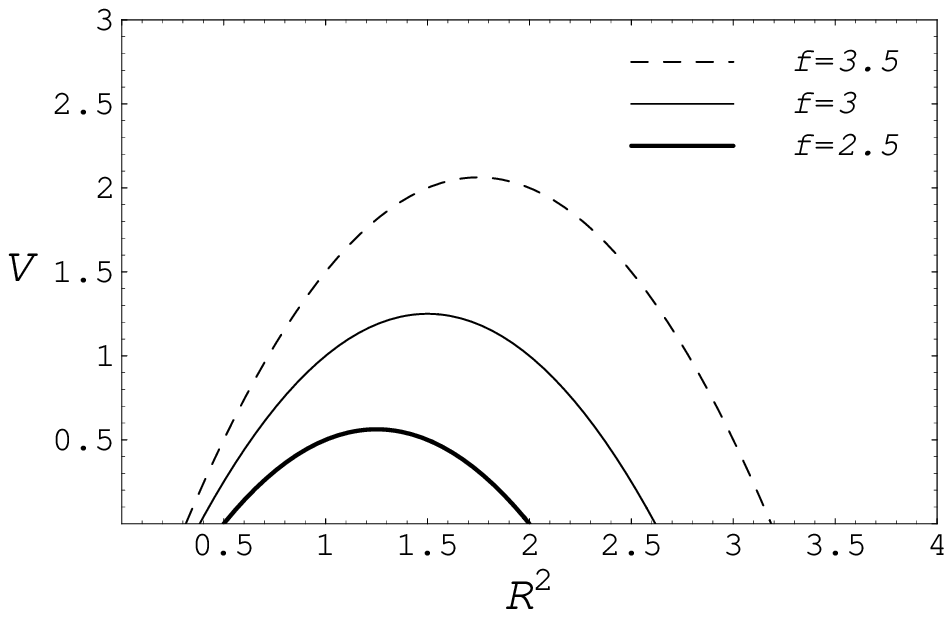,width=\hsize}
%\epsfxsize=1.0\hsize
%\epsfbox{plotd32a.eps}
%\label{fig:d32a}
\end{picture}\par
%\caption{Left}
\end{minipage}
\hfill
\begin{minipage}[t]{7.5cm}
\begin{picture}(7.5,5.0)
\epsfig{file=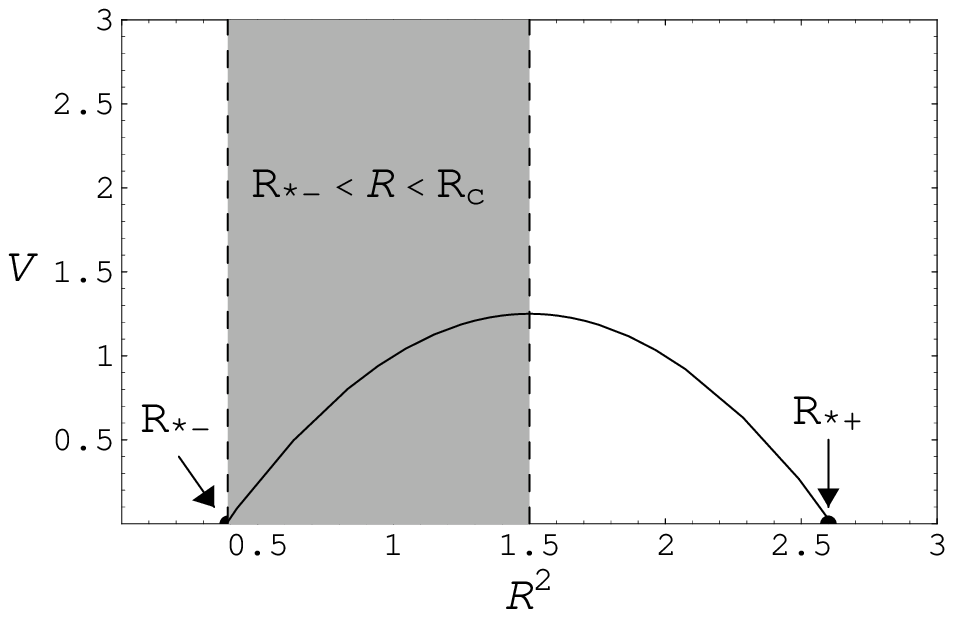,width=\hsize}
%\epsfxsize=1.0\hsize
%\epsfbox{plotd32a.eps}
%\label{fig:d32a}
\end{picture}\par
%\caption{Right}
\end{minipage}
\caption{\small Plots of $V$ for
$\Lambda<0$, $\beta>0$ and $Q>0$. Non-zero values of $f$ belong to the
interval $f>2\sqrt{-Q\Lambda/3}$ and correspond to shells of
constant $r$. In the lefthand plot ${R_c}<{R_{*-}}$ and gravity is
confined near the brane in ${R_{*-}}<R<{R_{*+}}$. In the
righthand plot ${R_{*-}}<{R_c}<{R_{*+}}$ and the shaded region indicates
where gravity is not bound to the brane.}
\label{fig:Vnpp}
\end{figure}

If $Q>0$
then $f>2\sqrt{-Q\Lambda/3}$ (see Fig.~\ref{fig:Vnpp}). We find two
regular rebouncing epochs at $R=R_{*\pm}$ where

\beq
{R_{*\pm}^2}=-{{3f}\over{2\Lambda}}\pm\sqrt{\beta}.
\eeq
The allowed phase space region is precisely the interval between these
two roots, ${R_{*-}}\leq R\leq{R_{*+}}$. Since the singularity at
${R_s}=0$ is inside the forbidden region it does not form and so we are in
the presence of globally regular solutions. In this case an initially
expanding dark radiation shell will rebounce at $R={R_{*+}}$ after
the time

\beq
{t_{*+}}(r)={1\over{2}}\sqrt{-{3\over{\Lambda}}}\left[{\pi\over{2}}-{\sin^{-1}}
\left({{{r^2}+{{3f}\over{2\Lambda}}}
\over{\sqrt{\beta}}}\right)\right].
\eeq
Then it reverses its motion to collapse until it reaches $R={R_{*-}}$
in the time

\beq
{t_{*-}}(r)={\pi\over{2}}\sqrt{-{3\over{\Lambda}}}.
\eeq
It rebounces again to expand to $R={R_{*+}}$ in the same amount
of time and subsequently repeats the
cycle. Thus these are oscillating globally regular solutions. Again note the
contrast with the dS scenario where no oscillating solutions
were found. Since $\Lambda<{\Lambda_c}$ there is only a gravity confinement
phase transition epoch in the cycle if ${R_{*-}}<{R_c}<{R_{*+}}$.

\subsection{Absence of cosmological constant: $\Lambda=0$}

When $\Lambda=0$ the potential is simply $V=f{R^2}-Q$. If $Q<0$ and
$f>0$ then there are no rebounce epochs in the allowed dynamical range
$R\geq 0$ and the singularity is at ${R_s}=0$ where $V(0,r)=-Q>0$. The
collapsing dark radiation shells reach the singularity
after the time

\beq
{t_s}(r)={1\over{f}}\left(\sqrt{f{r^2}-Q}-\sqrt{-Q}\right).
\eeq
In this case gravity never propagates out of the brane. If $Q>0$ and $f>0$ or
$Q<0$ and $-1<f<0$ there is a single rebounce
root at ${R_*}=\sqrt{Q/f}$. For $Q>0$ and $f>0$ the allowed dynamical
region is $R\geq{R_*}$ and so the singularity does not form at
${R_s}=0$. A collapsing shell falls to a minimum radius in
the time

\beq
{t_*}(r)={{\sqrt{f{r^2}-Q}}\over{f}}
\eeq
and then expands continuously to infinity. If ${R_*}>{R_c}$ gravity is
confined for all $R>{R_*}$. Otherwise the dark radiation goes
through a confinement phase transition at $R={R_c}$. For $Q<0$ and $-1<f<0$ the
dynamical range in phase space is $0\leq R\leq{R_*}$. The shells may
expand out to reach $R_*$ at

\beq
{t_*}(r)=-{{\sqrt{f{r^2}-Q}}\over{f}}.
\eeq
Then it reverses its motion to fall into the singular epoch ${R_s}=0$
in the proper time

\beq
{t_s}(r)=-{{\sqrt{-Q}}\over{f}}.
\eeq
In this setting gravity remains confined in the allowed dynamical
range, $0\leq R\leq{R_*}$.

\subsection{Absence of dark radiation: $Q=0$}

When the dark radiation degrees of freedom are not present $Q=0$ and
the potential to be considered is

\beq
V=V(R,r)={{\Lambda}\over{3}}{R^2}+f.
\eeq
For $\Lambda>0$ we may distinguish three types of evolution. If $-1<f<0$
there is a rebounce epoch towards infinite continuous expansion at
${R_*}=\sqrt{-3f/\Lambda}$ and the allowed dynamical region
is $R\geq{R_*}$. The time to collapse to minimum radius is

\beq
{t_*}(r)=\sqrt{{3\over{\Lambda}}}{\cosh^{-1}}
\left(r\sqrt{-{{\Lambda}\over{3f}}}\right).
\eeq
If $f>0$ there are no rebounce epochs for $R\geq 0$ and a singularity
forms at ${R_s}=0$ where the potential is $V(0,r)=f$. The time to meet
the singularity is

\beq
{t_s}(r)=\sqrt{{3\over{\Lambda}}}{\sinh^{-1}}
\left(r\sqrt{{\Lambda}\over{3f}}\right).
\eeq

If $\Lambda<0$ the only possibility is to take $f>0$. Then we find a
rebounce epoch at ${R_*}=\sqrt{-3f/\Lambda}$. After expansion to a
maximum radius $R_*$ in the time

\beq
{t_*}(r)=\sqrt{-{3\over{\Lambda}}}\left[{\pi\over{2}}-{\sin^{-1}}
\left(r\sqrt{-{{\Lambda}\over{3f}}}\right)\right].
\eeq
the brane collapses to the singularity at ${R_s}=0$ reaching it after
the time

\beq
{t_s}(r)={\pi\over{2}}\sqrt{-{3\over{\Lambda}}}.
\eeq
For $\Lambda<{\Lambda_c}$ the gravitational field is always
confined. Otherwise it never is.

\section{Conclusions}

We have analysed some aspects of the gravitational dynamics of inhomogeneous
dark radiation on a RS brane. This is an important first step towards the
understanding of a realistic collapse setting on a brane. Indeed, the
behaviour of dark radiation is nothing else than the dynamics of the vaccum
on the brane when the gravitational modes are excited only as energy density.

We have taken an effective 4-dimensional view point to show that with certain
simplifying, but natural assumptions, the Einstein field equations form a
solvable, closed system. The solutions obtained were shown to depend on the
brane cosmological constant $\Lambda$, the dark radiation tidal charge, $Q$,
and on the energy function $f(r)$.

We have given a precise description of the dynamics of the solutions and
characterized how this depends on $\Lambda$, $Q$ and $f(r)$. We have also presented
the conditions defining the solutions as singular or as globally regular.

Finally we have discussed the confinement of gravity to the vicinity of the
brane. We have seen that it depends on $\Lambda$, $Q$ and also on the brane
tension $\lambda$. If $\Lambda\leq{\Lambda_c}$ where ${\Lambda_c}={\tilde{\kappa}^4}
{\lambda^2}/12$ and $Q<0$ then the gravitational field is always confined to
the vicinity of the brane during its evolution. Alternatively, if $\Lambda<
{\Lambda_c}$ and $Q>0$ gravity is only bound to the brane for $R>{R_c}$ where
${R_c^4}=3Q/({\Lambda_c}-\Lambda)$. For $\Lambda>{\Lambda_c}$ and $Q<0$
confinement is restricted to the epochs $R<{R_c}$. We have noted that current
observations do not allow $\Lambda<0$ and $\Lambda\geq{\Lambda_c}$. Consequently,
the gravitational degrees of freedom may eventually propagate away from the
brane at high energies. Such a phase transition would occur at the epoch
$R={R_c}$ and should be detected in a change in Newton's gravitational constant.
\vspace{1cm}

\centerline{\bf Acknowledgements}

We thank the Funda\c {c}\~ao para a Ci\^encia e a Tecnologia (FCT) for
financial support under the contracts SFRH/BPD/7182/2001 and
POCTI/32694/FIS/2000. We also would like to thank Louis Witten and
T.P. Singh for helpful comments on the manuscript.


\begin{thebibliography}{30}

\bibitem{CLED}
J. Dai, R. G. Leigh and J. Polchinski, Mod.\ Phys.\ Lett.\ A {\bf 4},
2073 (1989);

R. G. Leigh, Mod.\ Phys.\ Lett.\ A {\bf 4}, 2767 (1989);

J. Polchinski, Phys.\ Rev.\ Lett. {\bf 75}, 4724 (1995)
\href{http://arXiv.org/abs/hep-th/9510017}{[arXiv:hep-th/9510017]};

P. Ho\v{r}ava and E. Witten, Nucl.\ Phys.\ {\bf B460}, 506 (1996)
\href{http://arXiv.org/abs/hep-th/9510209}{[arXiv:hep-th/9510209]};
Nucl.\ Phys.\ {\bf B475}, 94 (1996) \href{http://arXiv.org/abs/hep-th/9603142}
{[arXiv:hep-th/9603142]};
E. Witten, Nucl.\ Phys.\ {\bf B471}, 135 (1996)
\href{http://arXiv.org/abs/hep-th/9602070}{[arXiv:hep-th/9602070]};

T. Banks and M. Dine, Nucl.\ Phys.\ {\bf B479}, 173 (1996)
\href{http://arXiv.org/abs/hep-th/9605136}{[arXiv:hep-th/9605136]};

A. Lukas, B. A. Ovrut, K. S. Stelle and D. Waldram, Phys.\ Rev.\ D
{\bf 59}, 086001 (1999) \href{http://arXiv.org/abs/hep-th/9803235}{[arXiv:hep-th/9803235]};
Nucl.\ Phys.\ {\bf B552}, 246 (1999) \href{http://arXiv.org/abs/hep-th/9806051}
{[arXiv:hep-th/9806051]};

A. Lukas, B. A. Ovrut and D. Waldram, Phys.\ Rev.\ D {\bf 60}, 086001
(1999) \href{http://arXiv.org/abs/hep-th/9806022}{[arXiv:hep-th/9806022]};

N. Arkani-Hamed, S. Dimopoulos and G. Dvali, Phys.\ Lett.\ B {\bf 429},
263 (1998) \href{http://arXiv.org/abs/hep-ph/9803315}{[arXiv:hep-ph/9803315]};
Phys.\ Rev.\ D {\bf 59}, 086004 (1999)
\href{http://arXiv.org/abs/hep-ph/9807344}{[arxiv:hep-ph/9807344]};

K. R. Dienes, E. Dudas and T. Gherguetta, Phys.\ Lett.\ B {\bf 436}, 55
(1998) \href{http://arXiv.org/abs/hep-ph/9803466}{[arXiv:hep-ph/9803466]};
Nucl.\ Phys.\ {\bf B537}, 47 (1999) \href{http://arXiv.org/abs/hep-ph/9806292}
{[arXiv:hep-ph/9806292]};

I. Antoniadis, N. Arkani-Hamed, S. Dimopolous and G. Dvali,
Phys.\ Lett.\ B {\bf 436}, 257 (1998) \href{http://arXiv.org/abs/hep-ph/9804398}
{[arXiv:hep-ph/9804398]}.

\bibitem{LED}
K. Akama, Lect.\ Notes Phys.\ {\bf 176}, 267 (1982) \href{http://arXiv.org/abs/hep-th/0001113}
{[arXiv:hep-th/0001113]};

V. A. Rubakov and M. E. Shaposhnikov, Phys.\ Lett.\ B {\bf 125}, 139
(1983);

M. Visser, Phys.\ Lett.\ B {\bf 159}, 22 (1985) \href{http://arXiv.org/abs/hep-th/9910093}
{[arXiv:hep-th/9910093]};

E. J. Squires, Phys.\ Lett.\ B {\bf 167}, 286 (1986);

G. W. Gibbons and D. L. Wiltshire, Nucl.\ Phys.\ {\bf B287}, 117
(1987) \href{http://arXiv.org/abs/hep-th/0109093}{[arXiv:hep-th/0109093]}.

\bibitem{RS}
L. Randall and R. Sundrum, Phys.\ Rev.\ Lett. {\bf 83}, 3370 (1999)
\href{http://arXiv.org/abs/hep-ph/9905221}{[arXiv:hep-ph/9905221]};

L. Randall and R. Sundrum, Phys.\ Rev.\ Lett.\ {\bf 83}, 4690 (1999)
\href{http://arXiv.org/abs/hep-th/9906064}{[arXiv:hep-th/9906064]}.

\bibitem{GT}
J. Garriga and T. Tanaka, Phys.\ Rev.\ Lett. {\bf 84}, 2778 (2000)
\href{http://arXiv.org/abs/hep-th/9911055}{[arXiv:hep-th/9911055]}.

\bibitem{GW}
W. D. Goldberger and M. B. Wise, Phys.\ Lett.\ B {\bf 475}, 275 (2000)
\href{http://arXiv.org/abs/hep-ph/9911457}{[arXiv:hep-ph/9911457]}.

\bibitem{GKR}
S. Giddings, E. Katz and L. Randall, JHEP {\bf 03}, 023 (2000)
\href{http://arXiv.org/abs/hep-th/0002091}{[arXiv:hep-th/0002091]}.

\bibitem{SMS}
T. Shiromizu, K. I. Maeda and M. Sasaki, Phys.\ Rev.\ D {\bf 62}, 024012
(2000) \href{http://arXiv.org/abs/gr-qc/9910076}{[arXiv:gr-qc/9910076]};

M. Sasaki, T. Shiromizu and K. I. Maeda, Phys.\ Rev.\ D {\bf 62},
024008 (2000) \href{http://arXiv.org/abs/hep-th/9912233}{[arXiv:hep-th/9912233]}.

\bibitem{KR}
A. Karch and L. Randall, JHEP {\bf 0105}, 008 (2001)
\href{http://arXiv.org/abs/hep-th/0011156}{[arXiv:hep-th/0011156]}.

\bibitem{CEHS}
C. Cs\'aki, J. Ehrlich, T. J. Hollowood and Y. Shirman, Nucl.\ Phys.\
{\bf B581}, 309 (2000)
\href{http://arXiv.org/abs/hep-th/0001033}{[arXiv:hep-th/0001033]}.

\bibitem{CGC}
B. Carter, Phys.\ Rev.\ D {\bf 48}, 4835 (1993);

R. Capovilla and J. Guven, Phys.\ Rev.\ D {\bf 51}, 6736 (1995)
\href{http://arXiv.org/abs/gr-qc/9411060}{[arXiv:gr-qc/9411060]};

R. Capovilla and J. Guven, Phys.\ Rev.\ D {\bf 52}, 1072 (1995)
\href{http://arXiv.org/abs/gr-qc/9411061}{[arXiv:gr-qc/9411061]}.


\bibitem{BDL}
P. Bin\'etruy, C. Deffayet and D. Langlois, Nucl.\ Phys.\ {\bf B565}, 269
(2000) \href{http://arXiv.org/abs/hep-th/9905012}{[arXiv:hep-th/9905012]}.

\bibitem{FCS5D}
N. Kaloper, Phys.\ Rev.\ D {\bf 60}, 123506 (1999)
\href{http://arXiv.org/abs/hep-th/9905210}{[arXiv:hep-th/9905210]};

T. Nihei, Phys.\ Lett.\ B {\bf 465}, 81 (1999) \href{http://arXiv.org/abs/hep-ph/9905487}
{[arXiv:hep-ph/9905487]};

C. Cs\'aki, M. Graesser, C. Kolda and J. Terning, Phys.\ Lett.\ B {\bf
462}, 34 (1999) \href{http://arXiv.org/abs/hep-ph/9906513}{[arXiv:hep-ph/9906513]};

J. M. Cline, C. Grojean and G. Servant, Phys.\ Rev.\ Lett.\ {\bf 83},
4245 (1999) \href{http://arXiv.org/abs/hep-ph/9906523}{[arXiv:hep-ph/9906523]};

D. J. H. Chung and K. Freese, Phys.\ Rev.\ D {\bf 61}, 023511 (2000)
\href{http://arXiv.org/abs/hep-ph/9906542}{[arXiv:hep-ph/9906542]};

J. Lykken and L. Randall, JHEP {\bf 0006}, 014 (2000) \href{http://arXiv.org/abs/hep-th/9908076}
{[arXiv:hep-th/9908076]};

H. B. Kim and H. D. Kim, Phys. Rev.\ D\ {\bf 61}, 064003 (2000)
\href{http://arXiv.org/abs/hep-th/9909053}{[arXiv:hep-th/9909053]};

P. Kanti, I. I. Kogan, K. A. Olive and M. Pospelov, Phys.\ Lett.\ B {\bf
  468}, 31 (1999) \href{http://arXiv.org/abs/hep-ph/9909481}{[arXiv:hep-ph/9909481]};

P. Kraus, JHEP\ {\bf 9912}, 011 (1999) \href{http://arXiv.org/abs/hep-th/9910149}
{[arXiv:hep-th/9910149]};

\'E. \'E. Flanagan, S. H. Henry Tye and I. Wasserman, Phys.\ Rev.\ D {\bf 62},
044039 (2000) \href{http://arXiv.org/abs/hep-ph/9910498}{[arXiv:hep-ph/9910498]};

C. Cs\'aki, M. Graesser, L. Randall and J. Terning, Phys.\ Rev.\ D {\bf
  62}, 045015 (2000) \href{http://arXiv.org/abs/hep-ph/9911406}{[arXiv:hep-ph/9911406]}.

\bibitem{DR}
P. Bin\'etruy, C. Deffayet, U. Ellwanger and D. Langlois,
Phys.\ Lett.\ {\bf B477}, 285 (2000) \href{http://arXiv.org/abs/hep-th/9910219}
{[arXiv:hep-th/9910219]};

S. Mukohyama, Phys.\ Lett.\ B {\bf 473}, 241 (2000) \href{http://arXiv.org/abs/hep-th/9911165}
{[arXiv:hep-th/9911165]};

D. Ida, JHEP\ {\bf 0009}, 014 (2000) \href{http://arXiv.org/abs/gr-qc/9912002}
{[arXiv:gr-qc/9912002]};

S. Mukohyama, T. Shiromizu and K. I. Maeda, Phys.\ Rev.\ D {\bf 62},
024028 (2000) [Erratum-Phys.\ Rev.\ D {\bf 63}, 029901 (2001)]
\href{http://arXiv.org/abs/hep-th/9912287}{[arXiv:hep-th/9912287]};
P. Bowcock, C. Charmousis and R. Gregory, Class. Quant. Grav. {\bf
  17}, 4745 (2000) \href{http://arXiv.org/abs/hep-th/0007177}{[arXiv:hep-th/0007177]}.

\bibitem{SD}
P. Singh and N. Dadhich, {\it Localization of Gravity in Brane
  World Cosmologies}, \href{http://arXiv.org/abs/hep-th/0204190}{[arXiv:hep-th/0204190]}.

\bibitem{CHR}
A. Chamblin, S. W. Hawking and H. S. Reall, Phys.\ Rev.\ D {\bf 61}, 065007
(2000) \href{http://arXiv.org/abs/hep-th/9909205}{[arXiv:hep-th/9909205]}.

\bibitem{EHM}
R. Emparan, G. T. Horowitz and R. C. Myers, JHEP {\bf 0001}, 007
(2000) \href{http://arXiv.org/abs/hep-th/9911043}{[arXiv:hep-th/9911043]};

R. Emparan, G. T. Horowitz and R. C. Myers, JHEP {\bf 0001}, 021 (2000)
\href{http://arXiv.org/abs/hep-th/9912135}{[arXiv:hep-th/9912135]};

R. Emparan, G. T. Horowitz and R. C. Myers, Phys.\ Rev.\ Lett.\ {\bf
  85}, 499 (2000) \href{http://arXiv.org/abs/hep-th/0003118}{[arXiv:hep-th/0003118]}.

\bibitem{DMPR}
N. Dadhich, R. Maartens, P. Papadopoulos and V. Rezania,
    Phys.\ Lett.\ B {\bf 487}, 1 (2000) \href{http://arXiv.org/abs/hep-th/0003061}
    {[arXiv:hep-th/0003061]}.

\bibitem{ND}
N. Dadhich, Phys.\ Lett.\ B {\bf 492}, 397 (2000) \href{http://arXiv.org/abs/hep-th/0009178}
{[arXiv:hep-th/0009178]}.

\bibitem{GM}
C. Germani and R. Maartens, Phys.\ Rev.\ D {\bf 64}, 124010 (2001)
\href{http://arXiv.org/abs/hep-th/0107011}{[arXiv:hep-th/0107011]};

T. Wiseman, Phys.\ Rev.\ D {\bf 65}, 124007 (2002)
\href{http://arXiv.org/abs/hep-th/0111057}{[arXiv:hep-th/0111057]}; Class.\ Quant.\ Grav.\ {\bf 19},
3083 (2002) \href{http://arXiv.org/abs/hep-th/0201127}{[arXiv:hep-th/0201127]}.

\bibitem{DG}
N. Dadhich and S. G. Ghosh, Phys.\ Lett.\ B {\bf 518}, 1 (2001)
\href{http://arXiv.org/abs/hep-th/0101019}{[arXiv:hep-th/0101019]}.

\bibitem{BGM}
M. Bruni, C. Germani and R. Maartens, Phys.\ Rev.\ Lett.\ {\bf 87},
231302 (2001) \href{http://arXiv.org/abs/gr-qc/0108013}{[arXiv:gr-qc/0108013]}.

\bibitem{GD}
M. Govender and N. Dadhich, Phys.\ Lett.\ B {\bf 538}, 233 (2002)
\href{http://arXiv.org/abs/hep-th/0109086}{[arXiv:hep-th/0109086]}.

\bibitem{COSp}
J. Garriga and M. Sasaki, Phys.\ Rev.\ D {\bf 62}, 043523 (2000)
\href{http://arXiv.org/abs/hep-th/9912118}{[arXiv:hep-th/9912118]};

H. Kodama, A. Ishibashi and O. Seto, Phys.\ Rev.\ D {\bf 62}, 064022
(2000) \href{http://arXiv.org/abs/hep-th/0004160}{[arXiv:hep-th/0004160]};

D. Langlois, Phys.\ Rev.\ D {\bf 62}, 126012 (2000) \href{http://arXiv.org/abs/hep-th/0005025}
{[arXiv:hep-th/0005025]};

C. van de Bruck, M. Dorca, R. H. Brandenberger and A. Lukas, Phys.\ Rev.\
D {\bf 62}, 123515 (2000) \href{http://arXiv.org/abs/hep-th/0005032}{[arXiv:hep-th/0005032]};

K. Koyama and J. Soda, Phys.\ Rev.\ D {\bf 62}, 123502 (2000)
\href{http://arXiv.org/abs/hep-th/0005239}{[arXiv:hep-th/0005239]}.

\bibitem{MWBH}
R. Maartens, D. Wands, B. A. Bassett and I. P. C. Heard, Phys.\ Rev.\ D {\bf
  62}, 041301 (2000) \href{http://arXiv.org/abs/hep-ph/9912464}{[arXiv:hep-ph/9912464]}.

\bibitem{RM4dp}
R. Maartens, Phys.\ Rev.\ D {\bf 62}, 084023 (2000)
\href{http://arXiv.org/abs/hep-th/0004166}{[arXiv:hep-th/0004166]}.

\bibitem{LMSW}
D. Langlois, R. Maartens, M. Sasaki and D. Wands, Phys.\ Rev.\ D {\bf
  63}, 084009 (2001) \href{http://arXiv.org/abs/hep-th/0012044}{[arXiv:hep-th/0012044]}.

\bibitem{RM}
R. Maartens, {\it Geometry and Dynamics of the Brane World},\\
\href{http://arXiv.org/abs/gr-qc/0101059}{[arXiv:gr-qc/0101059]}.

\bibitem{CFT}
J. Maldacena, Adv.\ Theor.\ Math.\ Phys.\ {\bf 2}, 231 (1998)
\href{http://arXiv.org/abs/hep-th/9711200}{[arXiv:hep-th/9711200]};

S. S. Gubser, I. R. Klebanov and A. M. Polyakov,
Phys.\ Lett.\ B {\bf 428}, 105 (1998) \href{http://arXiv.org/abs/hep-th/9802109}
{[arXiv:hep-th/9802109]};

E. Witten, Adv.\ Theor.\ Math.\ Phys.\ {\bf 2}, 253 (1998)
\href{http://arXiv.org/abs/hep-th/9802150}{[arXiv:hep-th/9802150]};

O. Aharony, S. S. Gubser, J. Maldacena, H. Ooguri e
  Y. Oz, Phys.\ Rep.\ {\bf 323}, 183 (2000) \href{http://arXiv.org/abs/hep-th/9905111}
  {[arXiv:hep-th/9905111]}.

\bibitem{RSCFT}
S. S. Gubser, Phys.\ Rev.\ D {\bf 63}, 084017 (2001)
\href{http://arXiv.org/abs/hep-th/9912001}{[arXiv:hep-th/9912001]};

S. Nojiri, S. D. Odintsov and S. Zerbini, Phys.\ Rev.\ D {\bf 62},
064006 (2000) \href{http://arXiv.org/abs/hep-th/0001192}{[arXiv:hep-th/0001192]};

S. W. Hawking, T. Hertog and H. S. Reall, Phys.\ Rev.\ D {\bf 62},
043501 (2000) \href{http://arXiv.org/abs/hep-th/0003052}{[arXiv:hep-th/0003052]};

M. J. Duff and J. T. Liu, Phys.\ Rev.\ Lett.\ {\bf 85}, 2052 (2000)
[Class.\ Quant.\ Grav.\ {\bf 18}, 3207 (2001)]
\href{http://arXiv.org/abs/hep-th/0003237}{[arXiv:hep-th/0003237]};

S. Nojiri and S. D. Odintsov, Phys.\ Lett.\ B {\bf 484}, 119 (2000)
\href{http://arXiv.org/abs/hep-th/0004057}{[arXiv:hep-th/0004057]};

L. Anchordoqui, C. Nunez and K. Olsen, JHEP\ {\bf 0010}, 050 (2000)
\href{http://arXiv.org/abs/hep-th/0007064}{[arXiv:hep-th/0007064]};

S. de Haro, K. Skenderis and S. N. Solodukhin, Class.\ Quant.\ Grav.\
{\bf 18}, 3171 (2001) \href{http://arXiv.org/abs/hep-th/0011230}{[arXiv:hep-th/0011230]};

T. Shiromizu and D. Ida, Phys.\ Rev.\ D {\bf 64}, 044015 (2001),
\href{http://arXiv.org/abs/hep-th/0102035}{[arXiv:hep-th/0102035]}.

\bibitem{TPS}
T. P. Singh, Class.\ Quant.\ Grav.\  {\bf 16}, 3307 (1999)
\href{http://arXiv.org/abs/gr-qc/9808003}{[arXiv:gr-qc/9808003]}.

\bibitem{RW}
R. M. Wald, {\it{General Relativity}} (The University of
Chicago Press, Chicago, 1984).

\bibitem{qexp}
K. Ichiki, M. Yahiro, T. Kajino, M. Orito and G. J. Mathews,
{\it{Observational Constraints on Dark Radiation in Brane Cosmology}},
\href{http://arXiv.org/abs/astro-ph/0203272}{[arXiv:astro-ph/0203272]}.

\bibitem{EXP}
A. G. Riess {\it et al.}  [Supernova Search Team Collaboration],
Astron.\ J.\  {\bf 116}, 1009 (1998)
\href{http://arXiv.org/abs/astro-ph/9805201}{[arXiv:astro-ph/9805201]};
Astrophys.\ J.\  {\bf 560}, 49 (2001)
\href{http://arXiv.org/abs/astro-ph/0104455}{[arXiv:astro-ph/0104455]};

S. Perlmutter {\it et al.}  [Supernova Cosmology Project Collaboration],
Astrophys.\ J.\  {\bf 517}, 565 (1999)
\href{http://arXiv.org/abs/astro-ph/9812133}{[arXiv:astro-ph/9812133]};

S. Perlmutter, M. S. Turner and M. White,
Phys.\ Rev.\ Lett.\  {\bf 83}, 670 (1999)
\href{http://arXiv.org/abs/astro-ph/9901052}{[arXiv:astro-ph/9901052]};

N. A. Bahcall, J. P. Ostriker, S. Perlmutter and P. J. Steinhardt,
Science {\bf 284}, 1481 (1999) \href{http://arXiv.org/abs/astro-ph/9906463}
{[arXiv:astro-ph/9906463]};

U. Ellwanger, {\it{The Cosmological Constant}}, \href{http://arXiv.org/abs/hep-ph/0203252}
{[arXiv:hep-ph/0203252]}.

\bibitem{HE}
S. W. Hawking and G. F. R. Ellis, {\it{The Large Scale Structure of
    Space-time}} (Cambridge University Press, Cambridge, England, 1973).

\bibitem{PSJ}
P. S. Joshi, {\it{Global Aspects in Gravitation and Cosmology}}
(Clarendon Press, Oxford, England, 1993).

\bibitem{DJCJ}
S. S. Deshingkar, S. Jhingan, A. Chamorro and P. S. Joshi, Phys.\
Rev.\ D {\bf 63}, 124005 (2001) \href{http://arXiv.org/abs/gr-qc/0010027}{[arXiv:gr-qc/0010027]}.

\end{thebibliography}
\end{document}